\RequirePackage{fix-cm}
\documentclass[final]{svjour3}
\smartqed  
\usepackage{graphicx}
\usepackage{comment}
\usepackage[colorlinks,citecolor=blue,linkcolor=blue,urlcolor=blue]{hyperref}
\usepackage{breqn}
\usepackage{mathrsfs}
\usepackage{color}
\usepackage[normalem]{ulem}
\usepackage{cite}
\usepackage[left]{lineno}

\newcommand{\psat}{$P_{\mathrm{sat}}$ }
\newcommand{\beforesection}[1]{
 \vspace{-#1mm}
}
\newcommand*\patchAmsMathEnvironmentForLineno[1]{
  \expandafter\let\csname old#1\expandafter\endcsname\csname #1\endcsname
  \expandafter\let\csname oldend#1\expandafter\endcsname\csname end#1\endcsname
  \renewenvironment{#1}
     {\linenomath\csname old#1\endcsname}
     {\csname oldend#1\endcsname\endlinenomath}}
\newcommand*\patchBothAmsMathEnvironmentsForLineno[1]{
  \patchAmsMathEnvironmentForLineno{#1}
  \patchAmsMathEnvironmentForLineno{#1*}}
\AtBeginDocument{
\patchBothAmsMathEnvironmentsForLineno{equation}
\patchBothAmsMathEnvironmentsForLineno{align}
\patchBothAmsMathEnvironmentsForLineno{flalign}
\patchBothAmsMathEnvironmentsForLineno{alignat}
\patchBothAmsMathEnvironmentsForLineno{gather}
\patchBothAmsMathEnvironmentsForLineno{multline}
}

\newcommand{\Add}[1]{\textcolor{black}{#1}}
\newcommand{\AddGrammer}[1]{\textcolor{black}{#1}}
\newcommand{\Erase}[1]{\if0{#1}\fi}

\begin{document}

\title{Development of the characterization methods without electrothermal feedback
for TES bolometers for CMB measurements
}

\titlerunning{Development of the characterization methods for TES bolometers}        

\author{Yume Nishinomiya${}^{1}$ \and
Akito Kusaka${}^{1,2,3,4}$ \and Kenji Kiuchi${}^{1}$ \and Tomoki Terasaki${}^{1}$ \and Johannes Hubmayr${}^{5}$ \and Adrian Lee${}^{2,6}$ \and Heather McCarrick${}^{7}$ \and Aritoki Suzuki${}^{2}$ \and Benjamin Westbrook${}^{6}$
}

\authorrunning{Y. Nishinomiya et al.} 

\institute{Yume Nishinomiya \at
              \email{ynishinomiya@cmb.phys.s.u-tokyo.ac.jp}     \\      
            ${}^{1}$ Department of Physics, University of Tokyo, Tokyo 113-0033, Japan \\
            ${}^{2}$ Physics Division, Lawrence Berkeley National Laboratory, Berkeley, CA 94720, USA \\
            ${}^{3}$ Kavli Institute for the Physics and Mathematics of the Universe (WPI), Berkeley Satellite, 
            the University of California, Berkeley 94720, USA \\
            ${}^{4}$ Research Center for the Early Universe, School of Science, The University of Tokyo, Bunkyo-ku, Tokyo 113-0033, Japan \\
            ${}^{5}$ NIST Quantum Sensors Group, 325 Broadway Mailcode 687.08, Boulder, CO 80305, USA \\
            ${}^{6}$ Department of Physics, University of California, Berkeley, CA 94720, USA \\
            ${}^{7}$ Joseph Henry Laboratories of Physics, Jadwin Hall, Princeton University, Princeton, NJ 08544, USA \\
}

\date{Received: date / Accepted: date}

\maketitle
\begin{abstract}
Superconducting \Add{Transition-Edge Sensor} (TES) bolometers are used for cosmic microwave background (CMB) observations.
We 
\AddGrammer{used a testbed to evaluate the thermal performance of TES bolometers in regard to the}
saturation power $P_{\mathrm{sat}}$ and intrinsic thermal time constant $\tau_0$.
We developed an evaluation method that \AddGrammer{is} 
complementary to methods with electrothermal feedback. 
In our method, 
\Add{the antenna termination resistor of the bolometer is directly 
biased with DC or AC electric power}
to simulate optical power, 
and \AddGrammer{the} TES is biased with small power, which \AddGrammer{allows} 
$P_{\mathrm{sat}}$ and $\tau_0$ to be determined without contribution from the negative electrothermal feedback. 
We describe the method and results of the measurement using it. 
We evaluated $P_{\mathrm{sat}}$ of five samples by applying DC power 
and confirmed
\Add{the overall trend between $P_{\mathrm{sat}}$ and the inverse leg length}.
We evaluated $\tau_0$ of the samples by applying DC plus AC power, 
and the measured value was reasonable \AddGrammer{in consideration of} the expected values of other TES parameters. 
This evaluation method \AddGrammer{enables} 
us to verify whether \AddGrammer{a} TES \AddGrammer{has been} 
fabricated \AddGrammer{with} 
the designed values and to provide feedback for fabrication for future CMB observations. 

\end{abstract}


\section{Introduction}
\label{intro}
\beforesection{3}
\Add{A Transition-Edge Sensor (TES)~\cite{Irwin2005} 
has a narrow temperature region 
between normal and superconducting states and 
is used as a thermometer in a bolometer, 
which can measure the power of incoming photons.
TES bolometers have been adopted in cosmic microwave background (CMB) observations 
due to their supreme sensitivity~\cite{adachi2020,choi2020atacama,sayre2020,ade2021}}.

\Add{In this paper, we focus on evaluation of two relevant TES parameters:}
the saturation power \psat and intrinsic thermal time constant $\tau_0$. 
\Add{\psat is the amount of electrical power required to drive the TES normal in the absence of absorbed optical power.}
\Add{Its desired value is informed 
by the optical power from the sky absorbed by the TES bolometers~\cite{hill2018}}. 
\psat is related to TES thermal carrier noise 
due to fluctuations in heat flow between TES and the heat bath 
as $\mathrm{NEP}_{\mathrm{g}} \propto \sqrt{P_{\mathrm{sat}}}$, 
where $\mathrm{NEP}_{\mathrm{g}}$ is the noise equivalent power of the thermal carrier noise~\cite{Mather1982}. 
$\tau_0$ is defined as $C/G$, which is the ratio of the heat capacity of \AddGrammer{the}
bolometer island $C$ and the thermal conductivity $G$ between the TES
and \AddGrammer{the} heat bath. 
The response speed of TES is faster than 
$\tau_0$ due to negative electrothermal feedback. 
The time constant with the feedback 
can be defined
as $\tau_{\mathrm{eff}} = \tau_0 / (\mathscr{L}+1)$, 
where $\mathscr{L}$ is the loop gain related to \AddGrammer{the} TES bias power. 
Since $\tau_{\mathrm{eff}}$ acts as a low-pass filter for the TES current, 
the requirement of $\tau_{\mathrm{eff}}$ for CMB experiments 
is set \AddGrammer{based on} 
the sampling rate and signal modulation frequency~\cite{cothard2020} in addition to stability. 

\AddGrammer{We developed an evaluation method for \psat and $\tau_0$ without considering electrothermal feedback.
Our evaluation method enables us to provide feedback to the fabrication.
These parameters are tuned in the fabrication process \AddGrammer{of the TES}
to satisfy the observation requirements~\cite{westbrook2018}. 
In the fabrication,} \psat is tuned by varying the length of \AddGrammer{the} bolometer legs 
because it is proportional to $G$, \AddGrammer{while} 
$\tau_0$ is tuned by varying $C$. 
Our method is complementary to $I$-$V$ measurement with 
the feedback~\cite{stevens2020,cothard2020}, 
in which the TES current is measured \AddGrammer{while} sweeping 
\AddGrammer{the} bias voltage.
\Add{We can measure $\tau_0$ directly with our method, 
in contrast to the $I$-$V$ method where $\tau_0$ is estimated from 
$\tau_{\mathrm{eff}}$ as a function of the loop gain $\mathscr{L}$~\cite{cothard2020}.}
Here, we describe the methods and results of the measurement using them.

\beforesection{5}
\section{Saturation Power Measurement}
\label{Psat_test}
\beforesection{3}
\subsection{Setup}
\label{Psat_setup}
\begin{figure}[t]
\begin{minipage}{0.45\hsize}
\centering
\includegraphics[width=2.7cm,clip]{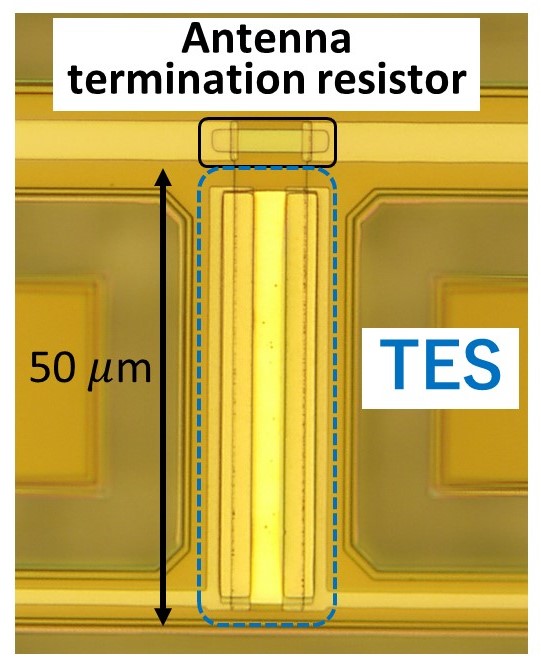}
\end{minipage}
\begin{minipage}{0.55\hsize}
\centering
\includegraphics[width=5.4cm,clip]{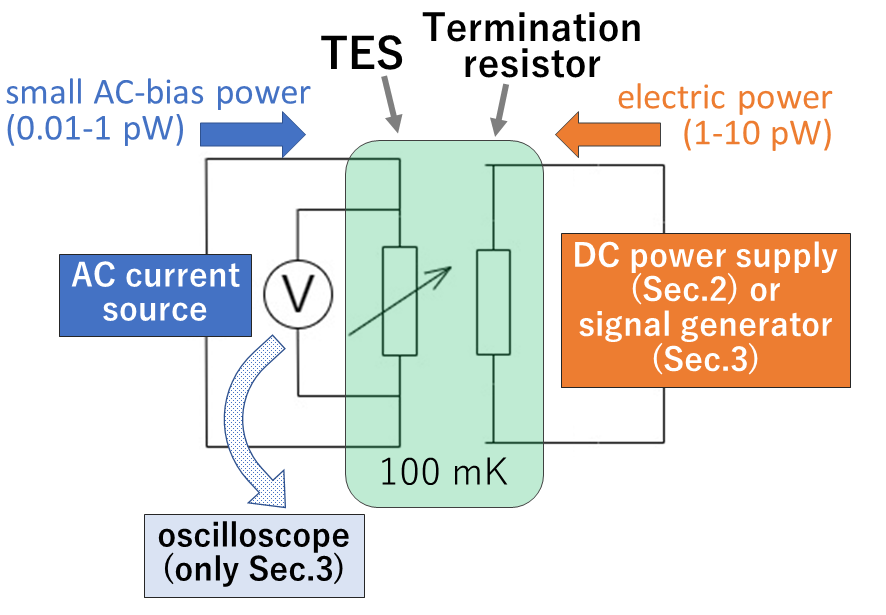}
\end{minipage}
\caption{Left: the photograph of the measured bolometer sample. 
The antenna termination resistor is \Add{thermally}
coupled with the TES. 
Right: the circuit of the measurement setup. 
TES was biased with small AC power. 
An oscilloscope was used only in the 
time constant measurement (Sec.~\ref{tau0_test}).
A DC power supply (Sec.~\ref{Psat_test}) or a signal generator (Sec.~\ref{tau0_test}) 
provided electric power to the termination resistor. 
(Color figure online)}
\label{fig:TES_sample}
\end{figure}

\beforesection{3}
\Add{We measured TES test samples for the development of a next-generation 
ground-based CMB experiment, provided by UC Berkeley.}
The samples have an antenna termination resistor coupled with a TES 
\Add{thermally}
as shown in \AddGrammer{Figure}~\ref{fig:TES_sample} (left). 
In actual operation, optical signals from the antenna travel on microstrip transmission lines, 
and the termination resistor converts the optical \AddGrammer{signals into} thermal signals. 
Here, instead we applied DC electric power to the termination resistor to simulate an optical signal. 

\AddGrammer{Figure} \ref{fig:TES_sample} (right) shows the circuit of the measurement setup. 
The bolometer sample 
was installed on the mixing chamber stage of a 
dilution refrigerator\footnote{Oxford Io, \url{https://nanoscience.oxinst.com/}}.
The stage temperature was controlled by a heater on the stage 
under closed-loop PID control with an accuracy of around 0.1~mK. 
Electric power to the termination resistor was 
provided by a room-temperature DC power supply.
The TES was biased with a small AC current with frequency of 9.8 Hz 
with an AC resistance bridge\footnote{LakeShore372, \url{https://www.lakeshore.com/}}. 
We measured the TES resistance by recording the TES voltage. 
The bias power ($\sim$ 0.01 pW) was much smaller than $P_{\mathrm{sat}}$ ($\sim$ 1-10~pW),
so $P_{\mathrm{sat}}$ could be determined without electrothermal feedback. 

\beforesection{5}
\subsection{Evaluation method and results}
\label{Psat_method}
\beforesection{2}
We defined \psat as equivalent to $P_{\mathrm{bath}}(T_{\mathrm{c}})$, 
which is the power flow to the heat bath 
when the TES island temperature is at critical temperature $T_{\mathrm{c}}$. 
\AddGrammer{This can be expressed using} 
the bath temperature $T_{\mathrm{b}}$, index $n$, and $G$ \cite{Irwin2005}: 
\beforesection{1}
\begin{equation}
P_{\mathrm{sat}} = P_{\mathrm{bath}}(T_{\mathrm{c}}) = 
K(T_{\mathrm{c}}^n - T_{\mathrm{b}}^n)
= \Add{\frac{G}{nT_{\mathrm{c}}^{n-1}}} (T_{\mathrm{c}}^n - T_{\mathrm{b}}^n)
\label{eq:Psat}
\end{equation}
where the thermal conductivity at $T_{\mathrm{c}}$ is well defined using constant $K$ as 
$G=\mathrm{d}P_{\mathrm{bath}}(T)/\mathrm{d}T |_{T_{\mathrm{c}}} = nKT_{\mathrm{c}}^{n-1}$. 
At steady state, $P_{\mathrm{bath}} = P_{\mathrm{opt}} + P_{\mathrm{bias}}$, 
where $P_{\mathrm{opt}}$ is the optical power TES receives,
and $P_{\mathrm{bias}}$ is the TES bias power.
In our measurements, $P_{\mathrm{opt}}$ was replaced with \AddGrammer{the} 
DC electric power dissipated in the resistor, $P_{\mathrm{res}}$ ($\sim$1-10~pW), 
and $P_{\mathrm{bias}}$ ($\sim$ 0.01~pW) could be neglected. 
Therefore, we assumed that 
$P_{\mathrm{bath}} = P_{\mathrm{res}}$. 

\begin{figure}[t]
\begin{minipage}{0.5\hsize}
\centering
\includegraphics[height=3.6cm]{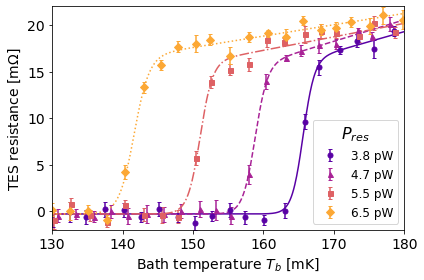}
\end{minipage}
\begin{minipage}{0.5\hsize}
\centering
\includegraphics[height=3.6cm]{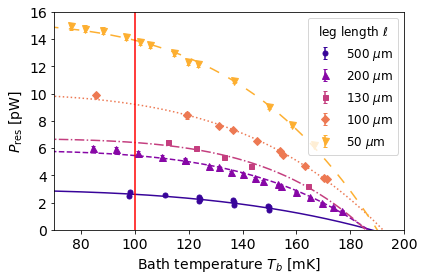}
\end{minipage}
\caption{Left: the measured resistance of one sample (100~$\mu$m leg length) as a function of
bath temperature for various $P_{\mathrm{res}}$.  
The error bars were calculated by the standard deviation from multiple measurements and the systematics.
Right: the relation between $P_{\mathrm{res}}$ and $T_{\mathrm{b,trans}}$ of five samples with different leg length $\ell$. 
The lines show fitting results with Eq.~\eqref{eq:Psat}.
\psat at 100~mK was estimated from the fitting lines
as the red vertical line shows.
(Color figure online)}
\label{fig:Psat_plot}
\end{figure}

First, we measured the TES resistance $R$ \AddGrammer{while} 
sweeping $T_{\mathrm{b}}$ without applying $P_{\mathrm{res}}$ 
and then determined $T_{\mathrm{c}}$ from the $R$-$T_{\mathrm{b}}$ 
curve\footnote{\Add{In $I$-$V$ measurement, $T_{\mathrm{c}}$ is estimated 
from a fit of the $P_{\mathrm{sat}}$-$T_{\mathrm{bath}}$ curves\cite{stevens2020}.}}.
\Add{The small bias power in our method enables the measurement of
$T_{\mathrm{c}}$ at zero current; increasing the current may shift and widen the transition~\cite{Bennett2012}.}
Next, we applied DC power to the termination resistor and repeated the $R$-$T_{\mathrm{b}}$ curve measurement. 
\AddGrammer{Figure}~\ref{fig:Psat_plot} (left) shows the result of the $R$-$T_{\mathrm{b}}$ curves of one bolometer for various $P_{\mathrm{res}}$.
\Add{To consider systematic errors, the error bars were re-scaled so that the reduced chi square reached 1.}
We found that the transition occurred at lower $T_{\mathrm{b}}$ when higher $P_{\mathrm{res}}$ was applied. 
From the $R$-$T_{\mathrm{b}}$ curves, we determined $T_{\mathrm{b,trans}}$, 
which is $T_{\mathrm{b}}$ in transition with a given $P_{\mathrm{res}}$. 
\AddGrammer{Figure}~\ref{fig:Psat_plot} (right) shows the relation 
between $P_{\mathrm{res}}$ and $T_{\mathrm{b,trans}}$ of five TES samples 
with different leg length. 
The data were fitted with Eq.~\eqref{eq:Psat} to estimate $G$ and $n$.  
Then we determined $P_{\mathrm{sat}}$ at \AddGrammer{the nominal operating temperature of} 100~mK.

\begin{figure}[t]
\centering
\includegraphics[height=3.5cm,clip]{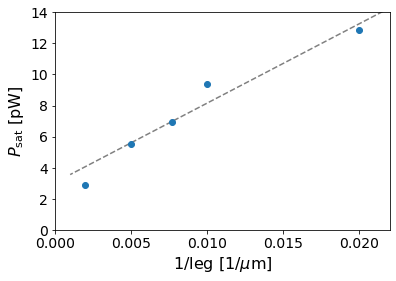} 
\caption{The measured $P_{\mathrm{sat}}$ as a function of the inverse leg length. 
The dashed line is a linear fit. (Color figure online)}
\label{fig:Psat_leg}
\end{figure}

The results of $G, n,$ and $P_{\mathrm{sat}}$ are shown in Table \ref{tab:thermal_fit}. 
All estimated $n$ values of the five samples were around 3,
\AddGrammer{which is reasonable} 
because $n$ is the thermal conductance index and was expected to be 
between 2 and 4 due to the material's properties~\cite{Mather1982}. 
$P_{\mathrm{sat}}$ was expected to be 
\Add{inversely proportional to the leg length}~\cite{suzuki2013multichroic,koopman2018}. 
As shown in \AddGrammer{Figure}~\ref{fig:Psat_leg},
we confirmed the overall trend between $P_{\mathrm{sat}}$ and the inverse leg length,
\Add{although the linear fit result was influenced by
possible systematic effects, such as the extra systematic errors in 
\psat fitting and slight geometrical variation of the five samples}.

\begin{table}[t]
\centering
\caption{The estimated parameters of five TES samples with different leg length $\ell$. 
\Add{The error of $T_{\mathrm{c}}$ comes from fitting of $R$-$T_{\mathrm{b}}$ curve.
The errors of $G$ and $n$ come from fitting of $P_{\mathrm{res}}$-$T_{\mathrm{b,trans}}$ curve. }
$P_{\mathrm{sat}}$ at 100~mK 
was determined using $G$ and $n$. 
}
\begin{tabular}{c c c c c c} \hline
ID & $\ell$ [$\mu$m] & $T_{\mathrm{c}}$ [mK] & $G$ [pW/K] & $n$ & $P_{\mathrm{sat}}$(100 mK) [pW] \\ \hline
1 & 500 & 187.7 $\pm$ 0.2 & 53 $\pm$ 4 & 2.9 $\pm$ 0.4 & 2.86 $\pm$ 0.06 \\ 
2 & 200 & 186.6 $\pm$ 0.2 & 103\Erase{.0} $\pm$ \Add{1} & 2.93 $\pm$ 0.03 & 5.51 $\pm$ \Add{0.04} \\
3 & 130 & 185.7 $\pm$ 0.2 & 129 $\pm$ 4 & 2.9 $\pm$ 0.2 & 6.9 $\pm$ 0.1 \\
4 & 100 & 192.1 $\pm$ 0.1 & 172 $\pm$ 1 & 3.03 $\pm$ 0.03 & 9.38 $\pm$ 0.03 \\
5 & 50 & 190.0 $\pm$ 0.3 & 248 $\pm$ 1 & 3.20 $\pm$ 0.03 & 12.86 $\pm$ 0.03 \\
\hline
\end{tabular}
\label{tab:thermal_fit}
\end{table}

\beforesection{4}
\section{Time Constant Measurement}
\label{tau0_test}
\beforesection{3}
\subsection{Setup}
\label{tau0_setup}
\beforesection{3}
\AddGrammer{To measure the intrinsic time constant $\tau_0$}, 
we applied AC electric power to the termination resistor, 
and measured the phase shift of the TES output voltage that was 
modulated by the electric power. 
The measurement setup was almost \Add{the} same as 
\AddGrammer{that of the} \psat measurement as described 
in Section~\ref{Psat_test}, except for two differences. 
Instead of a DC power supply, we used a signal generator 
to provide differential signals consisting of DC \AddGrammer{power} 
plus a small \AddGrammer{amount of} sinuous power. 
The AC resistance bridge can output a raw analog signal of TES voltage 
as well as a digital signal, so we measured the analog output signal using an oscilloscope. 

\beforesection{4}
\subsection{Evaluation method and results}
\label{tau0_method}
\beforesection{2}
\Add{The} TES thermal differential equation is shown below~\cite{Irwin2005}: 
\beforesection{1}
\begin{equation}
C \frac{\mathrm{d}T}{\mathrm{d}t} = -P_{\mathrm{bath}} + P_{\mathrm{bias}} + P_{\mathrm{opt}}
\label{eq:thermal}
\end{equation}
where $T$ is the TES temperature. 
We \AddGrammer{considered a} case \AddGrammer{where the} TES is in transition \AddGrammer{with a} small signal limit
\Add{and $P_{\mathrm{bath}} = P_{\mathrm{bath} 0} + G \Delta T $, 
where $P_{\mathrm{bath} 0}$ is $P_{\mathrm{bath}}$ at steady-state}.
\Add{In our measurement $P_{\mathrm{bias}}$ ($\sim$0.1 pW) is negligible.
$P_{\mathrm{opt}}$ is replaced with}
a DC offset plus small sinuous electric power with $f_{\mathrm{res}}$ applied to the termination resistor, 
\Add{$(P_{\mathrm{opt}}=)$}
$\,P_0 + \Delta P(t) = P_0 + P_{\mathrm{res}} \cos(2\pi f_{\mathrm{res}} t)$.
We can rewrite Eq.~\eqref{eq:thermal} and solve the differential equation 
\AddGrammer{for} $\Delta T(t)$:  
\beforesection{2}
\begin{equation}
\Delta T (t) = \frac{P_{\mathrm{res}}}{G} \left[\frac{1}{\sqrt{1+(2\pi f_{\mathrm{res}} \tau_0)^2}} \cos \left(2\pi f_{\mathrm{res}} t + \phi(f_{\mathrm{res}} ) \right)  \right]
\end{equation}
where
\beforesection{1}
\begin{equation}
\phi(f_{\mathrm{res}}) 
= - \arctan (2\pi f_{\mathrm{res}} \tau_0) 
\label{eq:phase2tau0}
\end{equation}
is the phase shift due to the electric power 
and $\tau_0=C/G$.
We can write \AddGrammer{the} TES resistance $R(t)$ 
\Add{as $R_{\mathrm{bias}}(1+\alpha \Delta T / T_{\mathrm{c}})$},
neglecting \AddGrammer{the} TES current term. 
When we bias TES with the AC current $I(t) = I_{\mathrm{bias}} \cos(2\pi f_{\mathrm{TES}}\, t)$, 
the TES output voltage $V=R(t)I(t)$ is: 
\begin{dmath}
V = R_{\mathrm{bias}} I_{\mathrm{bias}}
\left[ \cos(2\pi f_{\mathrm{TES}} \, t) + A 
\left( \cos (2\pi (f_{\mathrm{res}} + f_{\mathrm{TES}} )t + \phi ) 
+ \cos (2\pi (f_{\mathrm{res}} - f_{\mathrm{TES}} )t + \phi )\right)\right]
\label{eq:TESvolt}
\end{dmath}
\beforesection{1}
where $A= \alpha P_{\mathrm{res}}/(2 G T_{\mathrm{c}} \sqrt{1+(2\pi f_{\mathrm{res}} \tau_0)^2})$.
We measured $\phi$ from the TES output voltage (Eq.~\eqref{eq:TESvolt}) 
and estimated $\tau_0$ from Eq.~\eqref{eq:phase2tau0}. 

\begin{figure}[t]
\begin{minipage}{0.55\hsize}
\centering
\includegraphics[height=3.7cm,clip]{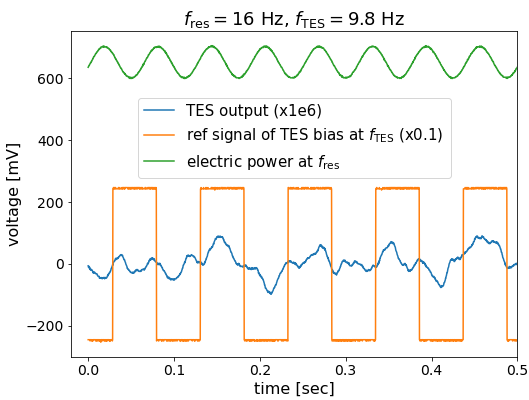}
\caption{The example of the time-ordered signals 
($f_{\mathrm{res}}=16$~Hz, $f_{\mathrm{TES}}=9.8$~Hz).
The blue line is the TES output signal amplified by $10^6$, 
orange is the reference square signal of $P_{\mathrm{bias}}$ at $f_{\mathrm{TES}}$, 
and green is the electric power including the DC offset and small sinuous wave at $f_{\mathrm{res}}$. 
(Color figure online)}
\label{fig:output_volt}
\end{minipage}
\hspace{5pt}
\begin{minipage}{0.45\hsize}
\centering
\includegraphics[height=3.7cm,clip]{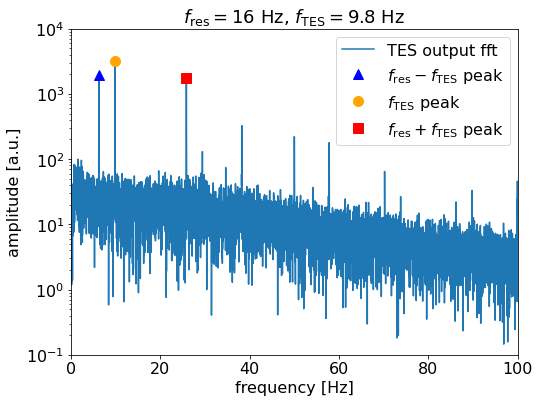}
\caption{The amplitude of the Fourier transform of TES output signal 
($f_{\mathrm{res}}=16$~Hz, $f_{\mathrm{TES}}=9.8$~Hz).
The square and triangle mark correspond to the components of frequency $(f_{\mathrm{res}} \pm f_{\mathrm{TES}})$. 
(Color figure online)}
\label{fig:TES_output_fft}
\end{minipage}
\end{figure}

\Add{We set $P_0 = $ 6.3 pW, 
which almost corresponds to the measured $P_{\mathrm{sat}}$, 
and $P_{\mathrm{res}} = $ 0.15 pW, which is $\sim$2\% of $P_{\mathrm{sat}}$.}
We swept $f_{\mathrm{res}}$ from 10 to 50 Hz for 4 hours, 
over which the bath temperature of $\sim$100 mK and the TES resistance on transition were kept stable.
We measured the TES voltage with an oscilloscope over 64 seconds 
for each $f_{\mathrm{res}}$ with a sampling rate of 3~kHz. 
\AddGrammer{Figure}~\ref{fig:output_volt} shows \AddGrammer{an} example of the time-ordered output signals 
with $f_{\mathrm{res}}$ of 16~Hz and $f_{\mathrm{TES}}$ of 9.8~Hz. 

\begin{figure}[t]
\begin{minipage}{0.6\hsize}
\centering
\includegraphics[height=3.7cm,clip]{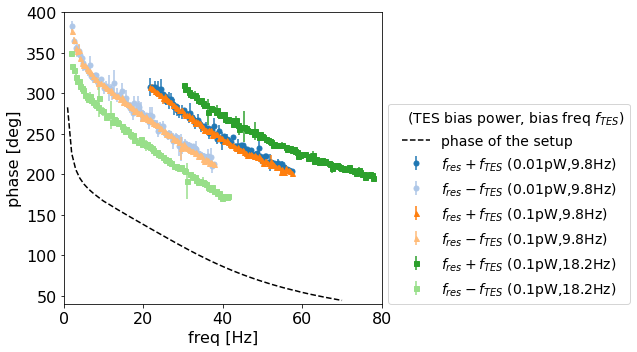}
\end{minipage}
\begin{minipage}{0.35\hsize}
\centering
\includegraphics[height=3.7cm,clip]{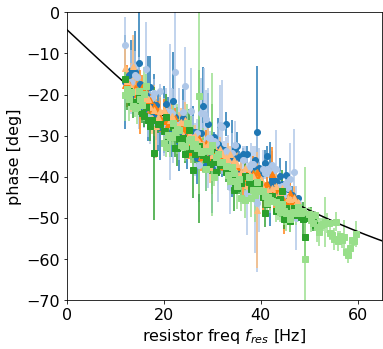}
\end{minipage}
\caption{Left: the phase of $f_{\mathrm{res}} \pm f_{\mathrm{TES}}$ components 
measured at various $P_{\mathrm{bias}}$ (0.01 or 0.1~pW) and $f_{\mathrm{TES}}$ (9.8 or 18.2~Hz).
Note that the x-axis is the actual frequency, not $f_{\mathrm{res}}$. 
The dashed line is the phase of the measurement setup. 
Right: the phase after subtraction of the phase of the measurement setup as a function of $f_{\mathrm{res}}$. 
The solid line is a fit result.
(Color figure online)}
\label{fig:measured_phase}
\end{figure}

The analysis procedure consisted of \AddGrammer{four} steps: 
(I) A template signal was made by mixing the resistor signal (Fig.~\ref{fig:output_volt} green)
with the TES input square signal, 
\Add{which had an amplitude of 5V TTL} 
(Fig. \ref{fig:output_volt} orange) in the time domain.  
The TES input square signal was provided by the AC resistance bridge and synchronized to $f_{\mathrm{TES}}$, 
so the template signal had two components proportional to 
$\cos ( 2\pi (f_{\mathrm{res}} \pm f_{\mathrm{TES}} )t)$. 
(II) To extract the information of the modulated waves from the TES output signal, 
$\cos (2\pi(f_{\mathrm{res}} \pm f_{\mathrm{TES}})t)$,
the Fourier transform of \AddGrammer{the} TES output signal 
(Fig.~\ref{fig:TES_output_fft}) was cross-correlated
with \AddGrammer{the} Fourier transform of the template from step (I). 
(I\hspace{-1pt}I\hspace{-1pt}I) The phase of the modulated waves was computed, and 
(I\hspace{-1pt}V) the phase delay due to the measurement setup, including the preamplifier and wirings, 
was subtracted as shown in \AddGrammer{Figure}~\ref{fig:measured_phase}. 

The phases are consistent 
after subtraction of that of the measurement setup in all measurement 
conditions (Fig.~\ref{fig:measured_phase} right), 
even though we measured the phase shift with different $f_{\mathrm{res}}$ 
and TES bias power.
The data were fitted with Eq.~\eqref{eq:phase2tau0} plus a DC offset $\phi_0$:
$\phi = - \arctan (2\pi f_{\mathrm{res}} \tau_0) + \phi_0$. 
$\tau_0$ of the TES sample was estimated as 3.0 $\pm$ 0.1~msec,
while $\phi_0$ was estimated as -4.6 $\pm$ 0.8 deg.
The systematic error of $\tau_0$ was estimated to be 0.6~msec, 
which is the difference of the fits between free $\phi_0$ and fixed $\phi_0=0$.
\Add{
The measured value of $\tau_0$ is consistent with the expected value within a factor of 2, 
considering the estimated value of $C$ 
(0.28 pJ/K) from the specific heat of the normal metal on the bolometer island~\cite{rayne1957} 
and the measured value of $G$ (129~pW/K) with the method described in Section~\ref{Psat_test}.}

\beforesection{5}
\section{Conclusion}
\label{conclusion}
\beforesection{3}
We demonstrated evaluation methods for TES parameters 
by applying DC or AC electric power to simulate optical power.
These methods enabled us to evaluate $P_{\mathrm{sat}}$ and 
$\tau_0$ without electrothermal feedback 
because only small TES bias power is applied.
These methods are necessary to characterize TES samples and to provide feedback for 
\AddGrammer{the} fabrication and development of TES bolometers for future CMB experiments~\cite{abazajian2016}.

\beforesection{2}
\begin{acknowledgements}
This work was supported by JSPS KAKENHI Grant Number
JP18H05539, JP19H04608, JP19K14736, JP19K21873.
Y.N. and T.T. acknowledge the support 
by FoPM, WINGS Program, the University of Tokyo.
\end{acknowledgements}

%
%

\bibliographystyle{spphys}       
\bibliography{references}

%
%

\end{document}